\documentclass{aa}
\usepackage{graphics}

\begin{document}

\thesaurus{08(08.04.1; 08.09.2 $\zeta$~Gem; 08.15.1; 08.22.1; 03.20.2)}

\title{The angular diameter and distance of the Cepheid $\zeta$~Gem}

\author{P. Kervella\inst{1} \and V. Coud\'e du Foresto\inst{2} \and G. Perrin\inst{2} \and M.
Sch\"oller\inst{1} \and W. A. Traub\inst{3} \and M. G. Lacasse\inst{3}}

\offprints{P. Kervella, pkervell@eso.org}

\institute{European Southern Observatory, Karl-Schwarzschildstr. 2, 85748 Garching, Germany
   \and{Observatoire de Paris-Meudon, 5, place Jules Janssen, 92195 Meudon Cedex, France}
   \and{Harvard-Smithsonian Center for Astrophysics, Cambridge, Massachusetts 02138, USA}} 

\date{Received 21 September 2000, Accepted 12 December 2000}

\maketitle

\begin{abstract}

   Cepheids are the primary distance indicators for extragalactic astronomy
			and therefore are of very high astrophysical interest.
   Unfortunately, they are rare stars, situated very far from Earth.
   Though they are supergiants, their typical angular diameter is only
   a few milliarcseconds, making them very challenging targets even
   for long-baseline interferometers. We report observations that were obtained in the
   $\rm K^{\prime}$ band (2-2.3 $\mu m$), on the Cepheid $\zeta$~Geminorum with the
   FLUOR beam combiner, installed at the IOTA interferometer.
   The mean uniform disk angular diameter was measured to be
   1.64~+0.14~-0.16~mas.
   Pulsational variations are not detected
   at a significant statistical level, but future observations with longer baselines should
   allow a much better estimation of their amplitude. The distance to $\zeta$~Gem
   is evaluated using Baade-Wesselink diameter
   determinations, giving a distance of 502~$\pm$~88~pc.


\end{abstract}


\section{Introduction \label{introduction}}

Cepheids are very important stars in the history of astronomy in general and
cosmology in particular. The relation discovered by H. Leavitt (Pickering \cite{pickering})
between the period of their light variation and their intrinsic luminosity makes
them essential tools for the estimation of large astronomical distances. As they
are intrinsically very bright stars, they can be detected in distant galaxies.
The Hubble Space Telescope has successfully observed cepheids in M100, and
measured a distance of 16.1 $\pm$ 1.8 Mpc (Ferrarese et al. \cite{ferrarese}).
The closest classical Cepheid ($\delta$~Cep) is situated at more than 300 pc.

Their resolution by ground based instruments is a challenge, as the angular
diameter of nearby Cepheids is typically smaller than 2 milliarcseconds (mas).
In order to determine precisely the zero point of the period-luminosity
relation, it is necessary to measure the angular diameter of a
number of Cepheids with high precision. The distance can then be derived from models of their
intrinsic diameter with much better precision than via direct parallax
measurement. The importance of Cepheid resolution by interferometry is
stressed in Sasselov \& Karovska (\cite{sasselov}).

Until recently, the only available technique to measure the angular diameter
of Cepheids was lunar occultation: it gave good results on $\zeta$~Gem
(Ridgway et al. \cite{ridgway}), with an angular size of 1.81~$\pm$~0.31~mas in K. The progress of
ground-based optical and infrared stellar interferometry allowed to measure a few Cepheid
angular diameters in the last years (Mourard et al. \cite{mourard}; Germain et al.
\cite{germain}). We report in this paper our observations of $\zeta$~Gem in the $\rm K^{\prime}$
band with the IOTA interferometer equipped with the FLUOR beam combiner.

In Sect.~\ref{section1}, we present the observations carried out at the IOTA interferometer
that resulted in the $\zeta$~Gem data. In Sect.~\ref{section2}, we interpret these
observations first in terms of mean angular diameter, and then we tentatively fit a
variable diameter model to our data. In Sect.~\ref{section3}, using the
extracted parameters, we can estimate by two different methods the distance to $\zeta$~Gem.
We first use a Baade-Wesselink (BW) estimation of the intrinsic diameter
of $\zeta$~Gem. Another promising technique to estimate directly the distance to
Cepheids is to use the angular amplitude of their pulsation coupled to radial
velocimetry. Finally, Sect.~\ref{section4} is dedicated to considerations for
future observations with the VLT Interferometer.


\section{Observations with FLUOR / IOTA \label{section1}}

\subsection{Presentation of FLUOR and IOTA}

IOTA (Infrared and Optical Telescope Array) is located on top of Mount Hopkins,
Arizona. It is a two telescope interferometer with 45~cm collecting apertures, and
selectable baselines from 5 to 38 meters. It is operated both at visible and
infrared wavelengths (Carleton et al. 1994; Traub 1998). A third telescope is
currently being installed, together with a second set of delay lines.

We have used the FLUOR (Fiber Linked Unit for Optical Recombination) beam
combiner. It is based on thin fluoride glass single-mode waveguides (core
diameter 6.5~$\mu$m) for observations between 2 and 2.3 $\mu$m ($\rm K^{\prime}$ band). A detailed
description of FLUOR can be found in Coud\'e du Foresto et al. (\cite{coude98}). The beam
combiner accepts the light from two telescopes and produces four output
signals: two photometric calibration signals (one for each beam) and two
complementary interferometric outputs. Both spatial filtering of the input
signals and beam combination are achieved simultaneously through the optical waveguides.
Atmospheric corrugation of the incoming wavefront is filtered out at the injection in
the fibers and converted into intensity fluctuations. The two photometric signals
allow us to calibrate these variations continuously.

This yields very accurate estimates of the squared modulus of the coherence factor $\mu^{2}$,
which is linked to the object visibility $V$ by the relationship
\begin{equation}
V^{2} = \mu^2 \ T^{2}
\end{equation}
where $T$ is the response of the system to a point source (interferometric efficiency).
$T$ is determined by bracketing the science target with observations of calibrator stars
whose $V$ is supposed to be known a priori.

The current accuracy on visibility estimates with FLUOR is approximately 1~\% for
most sources and is as good as 0.3~\% on bright objects.
 
\subsection{$\zeta$~Gem and its calibrators}

$\zeta$~Gem is the third brightest classical Cepheid in the northern sky
($\delta \ge$ -20 degrees) behind $\delta$~Cep and $\eta$~Aql.
Moreover, $\zeta$~Gem is the largest angular diameter Cepheid in
the northern sky. This star is therefore particularly interesting
for direct angular diameter measurements.

During observations, the interferometric efficiency
varies, which means that the science target observations have to be
calibrated periodically using observations of a known, stable star. The
calibration is then done by dividing the coherence factors obtained on the
target by the interferometric efficiency values, derived from the calibrator observations.
The choice of the calibrator is critical in the sense that any intrinsic
variation in the calibrator visibility will contaminate the science target final data.

\begin{table}
\caption{$\zeta$~Gem and calibrator parameters \label{parameters}}
\begin{tabular}{lccc}
\hline
Name & $\zeta$~Gem & HD 49968 & HD 62721\\
	& \object{HR 2650}	& \object{HR 2533} & \object{HR 3003}\\
$m_\mathrm{V}$	& 3.62-4.18	& 5.69 & 4.88\\
Mean $m_\mathrm{K}$ &	1.98 &	2.29 & 1.32\\
Spectral type	& F7Ib-G3Ib	& K5III & K4III\\
$T_{\mathrm{eff}}$	(K)& 5260-5780 &	3800 & 4000 \\
$\mathrm{Log}(g_{\mathrm{eff}})$&	1.5 &	2.5 & 2.5\\
$F_{B}$ ($\mu$m)& 13.11 & 13.21 & 13.21\\
Parallax (mas)	& 2.79 $\pm$ 0.81 &	6.36 $\pm$ 0.92 & 9.55 $\pm$ 0.83\\
$u$ & 0.282-0.241 & 0.379 & 0.370 \\
$\theta_{\rm {LD}}/\theta_{\rm {UD}}$ & 1.021-1.018 & 1.030 & 1.029 \\
$\theta_{\rm {UD}}$ (mas)	& $\sim$ 1.7	& 1.87 $\pm$ 0.02 & 2.93 $\pm$ 0.03 \\
Visibility &	$\sim$ 97~\%	& $\sim$ 96~\% & $\sim$ 93~\%\\
\hline
\end{tabular}
\end{table}

The selected calibrators, HD 49968 and HD 62721, are two bright K giants, located respectively 4.1
and 9.5 degrees away from $\zeta$~Gem (the parameters of all three stars
are presented in Table~\ref{parameters}). This proximity allows observations of both stars in
the same conditions, which limits the possible biases in the visibility
calibration to a minimum. Practically, it also allows us to reduce the pointing time and to stay in
the range of the delay line used to control the optical path difference.

The spectral type of these stars make them very unlikely variable stars, and they are not
reported variable in any catalogue of the SIMBAD database. Moreover, both calibrators have
been selected by Cohen et al. (\cite{cohen}) as stable reference objects for infrared
observations. These authors have derived a limb-darkened (LD) angular diameter of 1.93 $\pm$ 0.02
mas for HD 49968 and 3.02 $\pm$ 0.03 mas for HD 62721, based on high precision photometry at
infrared wavelengths. Our model fitting is done with a uniform disk model, therefore it is
necessary to convert these limb-darkened values to uniform disk (UD) equivalent diameters.

The difference between the limb-darkened and uniform disk diameters is given by
the relation (Hanbury-Brown et al. \cite{hanbury}):
\begin{equation}
\frac{\theta_{\rm {LD}}}{\theta_{\rm {UD}}} = \sqrt{\frac{1-u/3}{1-7u/15}}
\end{equation}
with $u$ = 0.379 and 0.370 the linear limb darkening coefficient in K (from Claret et al.
\cite{claret})  respectively for HD~49968 and HD~62721. This coefficient depends on the
effective temperature and gravity of the star (their values are listed in Table~\ref{parameters}).
This leads to a UD diameter of 1.87~$\pm$~0.02~mas for HD 49968, and of 2.93~$\pm$~0.03~mas
for HD~62721.  This value differs from the limb-darkened diameter by 3~\% , which is not
negligible compared to the expected angular diameter variations of $\zeta$~Gem (about 10~\%).
The $\zeta$~Gem limb darkening coefficient is also given for reference. It is variable as the
temperature of the star changes during the pulsation. For simplicity, a mean value of
$\theta_{\rm {LD}}/\theta_{\rm{UD}}$=1.020 is retained for the LD values given in the rest
of this paper.

Other limb related effects, such as possible $\zeta$~Gem limb brightening
(proposed by Sasselov \& Karovska \cite{sasselov}) are neglected. 

\subsection{Visibility measurements}

Due to the very small expected angular diameter of $\zeta$~Gem, we used the longest available
IOTA baseline, whose physical length is 38 meters.
The projected length on the sky during observations was about 37 meters. The precise
spatial frequencies sampled during the observations are indicated in Table~\ref{visibilities}
in cycles/arcsec.

The observations were obtained in March 1999, December 1999 and February 2000.
$\zeta$~Gem together with HD 49968 (during the 1999 campaigns) and HD 62721 (for February 2000)
were observed for sequences of twenty to thirty minutes alternatively. During each sequence,
100 to 600 interferograms were acquired in a row.

The data are acquired with FLUOR in scanning mode. A short stroke delay line (for the March 1999 campaign),
later replaced by a mirror mounted on a piezo stack (Ruilier \cite{ruilier}, installed in November 1999), sweeps
through the zero optical path difference position, while a Nicmos-III based infrared camera measures the
interferometric and photometric signals. For $\zeta$~Gem, the fringe frequency was 100 Hz. Dark current sequences
are recorded before and after each sequence and are used during the data reduction process for noise and signal
calibration. Sequences are acquired every few seconds (March 1999) or twice per second (December 1999 and
February 2000).

We have used the method developed by Coud\'e du Foresto et al. (\cite{coude97}) to reduce the data
and to derive the coherence factor modulus of the interference fringes from each batch of
observations. The interferometric efficiency of the instrument is computed from
the measurements obtained on the calibrators. This process and the determination of the object visibility are
described in details in Perrin et al. (\cite{perrin}). The calibrated visibilities measured on
$\zeta$~Gem are presented in Table~\ref{visibilities}.
A correction is applied to the calibrated visibilities to account for the difference in spectral
type between the calibrators and $\zeta$~Gem.
The corrective factors, called \textit{shape factors} ($F_{B}$), were computed for each star depending on its
spectral type (Chagnon \cite{chagnon}), and are listed in Table~\ref{parameters}.
They are of the order of 13 $\mu$m, and characterize the shape of the normalized spectral
intensity distribution:
\begin{equation}
F_{B} = \int_{0}^{+\infty}{B^{2}(\sigma) d\sigma} 
\end{equation}
with $B(\sigma)$ the spectral intensity distribution of the star normalized to unity:
\begin{equation}
\int_{0}^{+\infty}{B(\sigma) d\sigma} \equiv 1 
\end{equation}
The practical computation of $F_{B}$ is achieved by integrating the energy contained
in the observation spectral band ($\rm K^{\prime}$ in this paper) of standard stellar spectra
(taken from Cohen et al. \cite{cohen} and Wallace \& Hinkle \cite{wallace}).
Before the integration, the spectrum is degraded to the actual spectral resolution of FLUOR ($\sim 50$).
At such a low resolution, the spectral lines disappear. This explains the slow variability with the
spectral type (less than 1\% between $\zeta$~Gem and its calibrators).

\begin{table}
\caption{Calibrated visibilities obtained with FLUOR/IOTA on $\zeta$~Gem.}
\label{visibilities}
\begin{tabular}{llccr}
Julian & Phase & Calibrated & Sp.Freq. & Scans \\
Date -&  & Visibility & (cycles/ & \\
2451000 & & & arcsec) & \\
\hline
259.779 & 0.3491	& 0.9644	$\pm$ 0.0107	&	83.76 & 136 \\
262.722	&	0.6391 & 0.9777	$\pm$ 0.0108	&	84.25 & 157 \\
262.758	&	0.6426 & 0.9714	$\pm$ 0.0149	&	83.78 & 107 \\
595.838	&	0.4581 & 0.9932	$\pm$ 0.0085	&	83.84 & 316 \\
595.866 & 0.4609 & 0.9741 $\pm$ 0.0199 & 83.78 & 334 \\
602.734	&	0.1375 & 0.9700	$\pm$ 0.0116	&	85.82 & 396 \\
602.794	&	0.1435 & 0.9688	$\pm$ 0.0100	&	84.20 & 402 \\
\end{tabular}
\end{table}

\subsection{Data quality}

The data quality requirements for Cepheids observations are relatively higher than for
other programs, as the angular sizes of these stars are very
small, and their diameter variations are very subtle. Therefore, it is necessary to
select among the acquired data the best calibrated values to avoid the introduction of any bias in
the star parameters evaluation.

The data presented in this paper are the result of the selection of the highest quality
measurements over the three observing runs. The selection was done according to the
following rules:
\begin{itemize}
\item $\zeta$ Gem observations must be bracketed by a
calibrator observation, to reduce the risk of a biased visibility evaluation.
\item the difference between the two successive transfer function estimates (derived from the
bracketing calibrator observations) shall not be larger than the sum of the half-error bars
on those estimates:
\begin{equation}
\left| T_{i+1} - T_{i} \right| \le \sigma (T_{i}) + \sigma (T_{i+1})
\end{equation}
\item the ${\chi}^{2}$ between the calibrated visibility values from the two complementary
interferometric channels must be lower than 3. A large discrepancy between the two
channels would betray an instrumental or atmospheric perturbation.
\end{itemize}

Only 1848 out of the 7459 interferograms acquired on $\zeta$~Gem (25~\%) comply with all
the quality conditions, and only those resulting visibility measurements are presented in
Table~\ref{visibilities}. None of the data points obtained in December 1999 was qualified,
due to improper bracketing of $\zeta$~Gem by calibrators.


\section{$\zeta$~Gem angular diameter \label{section2}}

\subsection{Constant diameter model}

In this paragraph, we adopt a model of constant visibility for $\zeta$~Gem. We then compute
the mean angular diameter of this star over the observations.
This is done by applying a classical $ {\chi}^{2} $ minimization algorithm with
respect to the mean angular diameter only. The minimized quantity, relatively to the model
diameter is

\begin{equation}
\label{chi2sum}
{\chi}^{2} = \sum_{i}{\frac{(V_{\zeta}(\phi_{i}) -
V_{\rm model}(\phi_{i}))^2}{\sigma_{\zeta}(\phi_{i})^2}}
\end{equation}
with
\begin{equation}
V_{\rm model}(\phi_{i}) = \left|\frac{2 {\rm J_{1}}(z_{i})}{z_{i}}\right|
\end{equation}
\begin{equation}
z_{i} = \pi\ \theta_{\rm mean}\ f_{i}
\end{equation}

$\theta_{\rm mean}$ is the mean uniform disk diameter in arcseconds and $f_{\rm i}$ the
spatial frequency in cycles/arcsec for measurement $i$, at the phase $\phi_{i}$. The resulting
mean uniform disk diameter is presented in Table~\ref{results_constant}.

All the error bars in this paper are standard statistical plus or minus one $\sigma$
(standard deviation) error bars, yielding a probability of 68~\% for the extracted
parameters to be in the error bars assuming no systematic bias is present.

Nordgren et al. (\cite{nordgren}), observing with the Navy Prototype Optical Interferometer (NPOI),
find a UD diameter value of 1.48~$\pm$~0.08~mas (central wavelength $\lambda_{0}$=0.735 $\mu$m).
Independently, Lane et al. (\cite{lane}) measured a mean UD diameter of 1.65~$\pm$~0.3~mas with the Palomar Testbed
Interferometer (PTI), in the H band ($\lambda_{0}$=1.65 $\mu$m).
Our mean UD diameter value is consistent with these two recent measurements.

This estimate can also be compared with previous results from lunar occultation observations.
Ridgway et al. (\cite{ridgway}) have measured the $\zeta$~Gem diameter in
the J and K bands. They have obtained uniform disk diameters of 1.81~$\pm$~0.31~mas in K and
1.88~$\pm$~0.86~mas in J. Chandrasekhar (\cite{chandra}) has measured a maximum value of
1.9~$\pm$~0.3~mas in the K band. Ashok et al. (\cite{ashok}) obtained a value of
1.6~$\pm$~0.5~mas, also in K. These measurements and ours are compatible within the error
bars though our value seems to be slightly smaller than the previous measurements.

\begin{figure}[t]
\resizebox{\hsize}{!}{\includegraphics{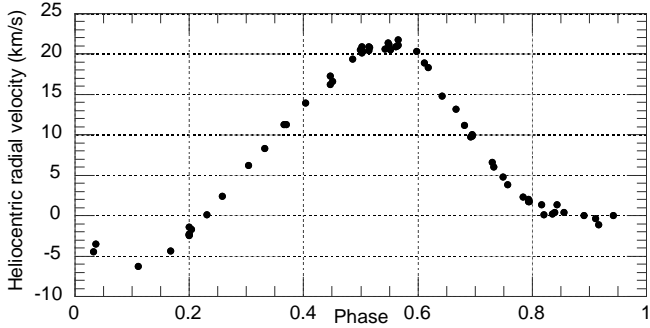}}
\caption{$\zeta$~Gem radial velocity, taken from Bersier et al. (\cite{bersier94}).
The pulsation period is 10.150079 days. \label{radvel}} 
\end{figure}

\begin{table}
\caption{Constant diameter model fit results \label{results_constant}}
\begin{tabular}{lcc}
Mean $\theta_{\rm {UD}}$ & 1.66 +0.14 -0.16 mas \\
Mean $\theta_{\rm {LD}}$ & 1.69 +0.14 -0.16 mas\\
Total ${\chi}^{2}$ & 6.08 \\
Reduced ${\chi}^{2}$ & 1.01\\
\end{tabular}
\end{table}

\subsection{Variable diameter model \label{variable_diameter}}

\subsubsection{Diameter variation curve from radial velocimetry data}

\begin{figure}[t]
  \resizebox{\hsize}{!}{\includegraphics{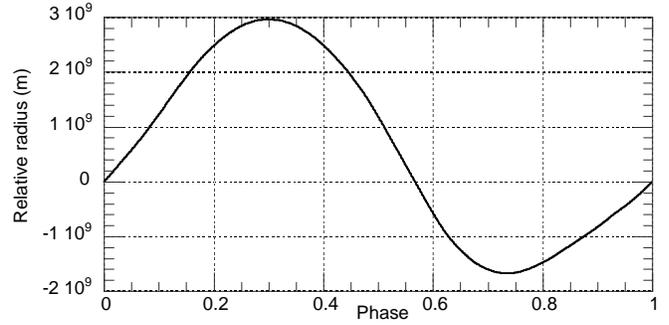}}
  \caption{Diameter variations of $\zeta$~Gem as integrated from the radial velocity data
  presented on Fig.~\ref{radvel}, using a $p$-factor of 1.36. These variations are given relatively
  to the star radius at phase zero, when the maximum luminosity of the star occurs (for visible
light).
  \label{radius}}
  \label{diamvar}
\end{figure}

A more realistic model for $\zeta$~Gem takes into account the angular diameter
variations as they are measured by spectroscopic radial velocimetry. In order to compute the
curve of the radius variation of $\zeta$~Gem, we integrated the high precision
($\sigma({\rm v})\le	$ 0.5 km/s) radial velocity measurements obtained with
the CORAVEL spectrograph by Bersier et al. (\cite{bersier94}), presented in Fig.~\ref{radvel}.
These measurements were phased using a period P = 10.150079 days and a reference
epoch $T_{0}$ = 2444932.736 taken from the same authors. The mean radial velocity of
$\zeta$~Gem (5.83 km/s) was subtracted from the radial velocity values before
integration.

The integration of the radial velocity curve requires to take into account the
limb darkening of the star. The limb darkening gives a higher relative
weight to the center of the disk than to the outer part. This means that the
apparent radial velocity is lower than the real pulsational velocity
value. The correction for this effect is included in a multiplicative term called the
projection factor (or $p$-factor). 

The $p$-factor can be derived using models of Cepheid atmospheres (Sabbey et al. \cite{sabbey})
or using the measured widths and asymmetries of spectral lines.
Its precise estimation is still an open question. New interferometers under
construction (see Sect.~\ref{section4}) will enable direct measurements of the limb
darkening on the closest Cepheids and hopefully will clarify this point.
This coefficient is essential in the distance
determination by the BW method (see Sect.~\ref{section41}). The most widely accepted
value for the $p$-factor of Cepheids is 1.36, and it has been used for our integration
of the CORAVEL data.

The resulting diameter curve is presented in Fig.~\ref{diamvar}. The shape of this
curve is typical of an intermediate period Cepheid star, with a maximum at
phase 0.3 and a minimum diameter at phase 0.7. For comparison, examples of typical
radii curves for two Cepheids: U Sgr, P=6.7 days and SZ Aql, P=17.1
days can be found in Laney et al. (1995).

\subsubsection{$\zeta$~Gem parameters}

We adjust a model with two parameters (the mean angular diameter and the amplitude
of the pulsation) to our visibility data. The shape of the expected angular diameter
variation curve is shown in Fig.~\ref{diamvar}. The fit is done in the visibility
space. We use the same classical ${\chi}^{2}$ minimization algorithm as for the constant
diameter model, with two parameters.

To extract the parameters of the $\zeta$~Gem diameter variations, the minimized
expression is the same as in Eq.~\ref{chi2sum}, but the model visibility is now:

\begin{equation}
V_{\rm model}(\phi_{i}) = \left|\frac{2 {\rm J_1}(z(\phi_{i}))}{z(\phi_{i})}\right| 
\end{equation}
with
\begin{equation}
z(\phi_{i}) = \pi \left(\theta_{\rm mean} + {\frac{\Delta D(\phi_{i})}{\Delta D_{\rm
max}}} A\right)  f_{i} 
\end{equation}
where:

\begin{itemize}
\item $ \theta_{\rm mean}$ is the mean diameter of the star. This is the first parameter
computed via the  ${\chi}^2$ fit.
\item $A$ is the pulsation amplitude, the second parameter adjusted by the  ${\chi}^2$ fit.
\item $\frac{\Delta D(\phi_{i})}{\Delta D_{\rm max}}$ is the normalized diameter variation
at phase $\phi_{i}$ as given by the integration of the radial velocity curve. It is normalized
to the total amplitude $\Delta D_{\rm max}$.
\item $f_{i}$ is the spatial frequency in cycles/arcsec for measurement $i$.
\end{itemize}

\begin{figure}
\resizebox{\hsize}{!}{\includegraphics{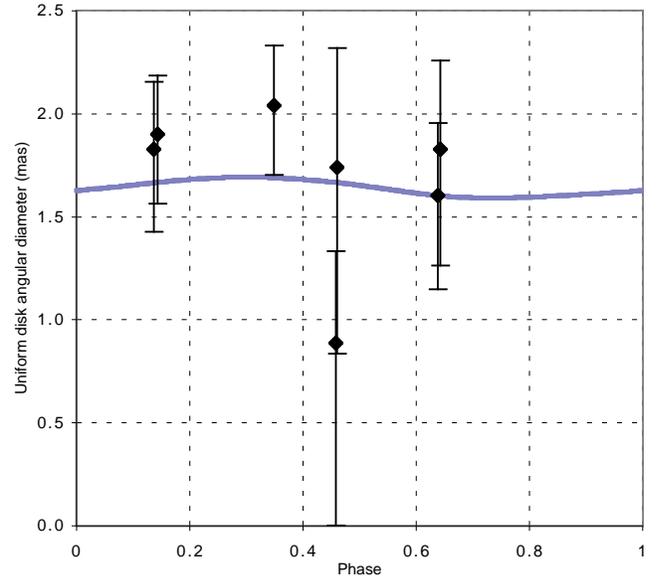}}
\caption{$\zeta$~Gem variable angular diameter model fit. The curve corresponds to the
parameters given in Table~\ref{results_variable}. \label{variable_fit}} 
\end{figure}

A plot of the result of the fit is presented in Fig.~\ref{variable_fit}. The parameters which
minimize the $ {\chi}^2 $ are listed in Table~\ref{results_variable}.

\begin{table}
\caption{Variable diameter model fit results \label{results_variable}}
\begin{tabular}{lcc}
Mean $\theta_{\rm {UD}}$ & 1.64 +0.14 -0.16 mas \\
Mean $\theta_{\rm {LD}}$ & 1.67 +0.14 -0.16 mas \\
Amplitude & 0.1 $\pm$ 0.3 mas \\
Total ${\chi}^{2}$ & 6.03 \\
Reduced ${\chi}^{2}$ & 1.21 \\
\end{tabular}
\end{table}

\subsection{Discussion}

The reduced ${\chi}^{2}$ value for the variable model fit is 20~\% larger than the value
obtained for the constant diameter model.

From this result we conclude that given the limited baseline of IOTA,
the detection of the diameter variations
is beyond the capabilities of our instrument. However, considering the
level of uncertainty on the individual points, both the constant and variable angular
diameter models are compatible with our results.


\section{Distance to $\zeta$~Gem \label{section3}}

\subsection{Using the Baade-Wesselink diameter estimate \label{section41}}

Following the suggestion from Sasselov \& Karovska (1994), we compute the distance to
$\zeta$~Gem by combining a mean radius from a complete B-W solution and our measurement of
this star's angular diameter. 

The Baade-Wesselink (BW) method assumes that we can observe simultaneously the emitting
surface of the star in flux, temperature variation and radial motion (Gautschy \cite{gautschy87}). The
flux and temperature variations give the ratio of the instantaneous radius of the star to
the mean radius through the equation (assuming $T_{\rm eff}$ is constant for simplicity):

\begin{equation}
{\left[ \frac{R(t_2)}{R(t_1)} \right]}^2 = \frac{L(t_2)}{L(t_1)} = 10^{-0.2 . (m(t_2)-m(t_1))}
\end{equation}

The integration of the radial velocity curve gives the linear amplitude
of the pulsation, making it possible to evaluate the radius itself.

Numerous BW measurements have been carried out on $\zeta$~Gem. A survey of eight
BW diameter measurements obtained before 1982 can be found in Fernie 
(\cite{fernie}). They range from 64 to 73~$D_{\odot}$. More recently, Krockenberger et al.
(\cite{krockenberger}), estimated the diameter of $\zeta$~Gem to be 69.1~+5.5~-4.8~$D_{\odot}$.
Bersier et al. (\cite{bersier97}) have derived a value of 89.5~$\pm$~13.3~$D_{\odot}$. For
coherence with the radial velocity measurements, this last value is retained in this paper's
computations. 

Other methods, derived from BW, have also been used, such as 'CORS' by Ripepi et al.
(\cite{ripepi}), yielding a diameter of 73.5~$D_{\odot}$, or 'new CORS', giving
86.2~$D_{\odot}$. These authors have also experimented with variable values of the $p$-factor on
$\zeta$~Gem during the pulsation, giving values of 80.1~$D_{\odot}$ (CORS) or 93.7~$D_{\odot}$
(new CORS).

Knowing the mean uniform disk angular diameter of $\zeta$~Gem and its intrinsic BW mean
diameter, 89.5~$\pm$~13.3~$D_{\odot} $ (from Bersier et al. \cite{bersier97}), it is now easy
to compute its distance:
\begin{equation}
\label{distance_eq}
d = \frac{9.305 \ D}{\theta_{\rm {UD}}}
\end{equation}
with $d$ the distance in parsec, $D$ the diameter in $D_{\odot}$, and $\theta_{\rm {UD}}$ the
angular diameter in mas.

The distance $d$ to $\zeta$~Gem can be derived knowing the constant diameter model $\theta_{\rm
{UD}}$ = 1.66~+0.14~-0.16~mas. Eq.~\ref{distance_eq} gives $d$($\zeta$~Gem) =
502~$\pm$~88~pc, equivalent to a parallax of 1.99~$\pm$~0.36 mas. The parallax obtained here
for $\zeta$~Gem is consistent with the parallax measured by Hipparcos : 2.79~$\pm$~0.81~mas,
though on the lower end of the error bars.

It is interesting to note that the error in the derived distance is currently dominated by the
uncertainty on the assumed BW mean diameter. Moreover, there is clearly a possible systematic
error at least as large as the reported statistical error, considering the wide range of discrepant
BW diameter values (64-94~$D_{\odot}$). The estimation of these systematics is beyond the scope
of our paper and the error bars given here do not include them.

\subsection{Method for distance determination by the amplitude of the pulsation \label{method_amplitude}}

\subsubsection{Principle}

It is also possible to estimate the distance to a pulsating star directly from the
geometrical amplitude of its pulsation. On one hand, we have the absolute amplitude of
the pulsation by integrating the radial velocimetry curve. On the other hand, we can measure
directly the angular diameter variation, though the amplitude determined in this paper is
affected by a large uncertainty. By combining these two values, one can derive the
distance to the star via the equation:
\begin{equation}
d = \frac{6.686 . 10^{-9} {\Delta D}}{\Delta \theta_{\rm {UD}}}
\end{equation}
with $d$ in parsec, $\Delta \theta_{\rm {UD}}$ in mas and $\Delta D$ in meters.

Intuitively, we use what we know along the line of sight (radial velocimetry gives a
linear amplitude value) and perpendicular to the line of sight (interferometry provides an
angular amplitude) to derive the distance. This assumes that $\zeta$~Gem
is essentially pulsating radially (see Gautschy \& Saio \cite{gautschy95} and
Gautschy \& Saio \cite{gautschy96} for an extensive review of star pulsations), but this hypothesis is
nowadays widely accepted (Bono et al. \cite{bono}).

\subsubsection{Visible diameter curve vs. infrared interferometric observations \label{integ_curve}}

While the radial velocity measurements are done mostly in the visible, the
interferometric observations are often done at infrared wavelengths (Lane et al. \cite{lane},
this paper). To assess the validity of the visible radius curve of the star to fit
infrared wavelengths interferometric data, it is necessary to compare what is really observed 
by interferometry at $2.2~\mu$m to what is measured by radial velocimetry
in the visible.

The formation of the absorption lines used to measure the radial velocity
through the Doppler shift happens at
different depths in the stellar atmosphere depending on the wavelength:
the longer the wavelength, the higher the line-forming region in the atmosphere. In the case
of the infrared, the lines are formed \textit{above} the visible ones.
In the absence of shock waves, it is therefore expected to see a larger pulsation amplitude
in the infrared than in the visible, as the infrared continuum-forming region
is located above the visible photosphere.

Regarding this question, Sasselov \& Lester (\cite{sasselov90})
have obtained measurements at $1.08$ and $1.6~\mu$m of the radial velocity of several Cepheids, and find
for $\zeta$~Gem an apparent velocity difference of 2.7~km/s at phase 0.802,
shortly after the minimum diameter. Confirming this trend, Butler et al. (\cite{butler}) 
have studied the cases of three bright Cepheids ($\delta$ Cep, P=5 days; $\eta$ Aql,
P=7 days; X Cyg, P=16 days). They obtained radial velocity curves from spectral lines
located in the near infrared ($\lambda =0.8~\mu$m) and in the visible
($\lambda =0.5~\mu$m). For the two shorter period Cepheids, the near infrared 
and visible radial velocity curves differ by less than 1~km/s, but the longer period
Cepheid X Cyg shows peak velocity differences of as much as 3~km/s.

Assuming a radius for $\zeta$~Gem of 89.5 $D_{\odot}$ ($1.25~10^{11}$~m)
taken from Bersier et al. (\cite{bersier97}), a 3 km/s radial velocity difference integrated
over half the period would correspond to 2~\% of the diameter of the star.
The effect of this difference on the diameter curve is that the amplitude is increased by
20\% (compared to the visible curve). Such a bias would also reduce accordingly the
apparent distance estimation through the amplitude of the pulsation.

Considering the fact that we do not detect the pulsations significantly, this effect is neglected
in Sect.~\ref{zeta_puls_dist}. However, when very high precision Cepheid diameter measurements will
be possible with instruments such as the VLT Interferometer (Glindemann et al. \cite{glindemann},
see Sect.~\ref{section4}),
it will be important to rely on radial velocimetry and interferometric observations
obtained at the same wavelengths for distance measurements.

In the case of longer period, large amplitude Cepheids, the suspected presence of
strong shock waves and velocity gradients (Butler et al. \cite{butler96})
make it more difficult to estimate the true infrared
pulsational velocity. Shorter period Cepheids are likely to be less affected, showing
smoother, more sinusoidal radial velocity curves than longer period stars.

\subsubsection{Application to $\zeta$~Gem \label{zeta_puls_dist}}

The amplitude of the pulsation as derived from the model fit presented in Sect.~\ref{variable_diameter}
is 0.1 $\pm$ 0.3 mas. The amplitude of the radius variation is
$4.64~10^{9}$~m (see Fig.~\ref{radius}), assuming that the atmosphere of the star is comoving
(same behavior between visible and infrared wavelengths). We obtain d($\zeta$~Gem) =
310~+$\infty$~-230~pc.

This distance is compatible with the Hipparcos measurement ($\pi$ = 2.79~$\pm$~0.81 mas,
equivalent to a distance of 358~+147~-80~pc).
Though this method is promising, the estimation of the distance to $\zeta$~Gem by the pulsation
amplitude is still beyond the present capabilities of our instrument. Moreover, due to the use of
visible radial velocity curves, our distance (as well as the one derived by Lane et al. \cite{lane})
may be biased by up to 20\% towards smaller values (see Sect. \ref{integ_curve} for discussion).


\section{Future observations with the VLT Interferometer \label{section4}}

In the next years, the VLT Interferometer (VLTI), CHARA and the Keck Interferometer will be
the instruments providing the highest angular resolution. It is
expected that their hundreds of meters baselines will allow the resolution of
the closest Cepheids. The VLT Interferometer (up to 200 meter baseline,
fringe tracking, large apertures with adaptive optics) will allow precise
diameter measurements on a large number of Cepheids, covering a wide range of
periods and distances. Both radial pulsations and non radial oscillations
will be measurable through interferometric observations by using different
baseline orientations.

Even assuming the same performances that were obtained with FLUOR (conservative 1~\%
visibility accuracy in 100 scans, without adaptive optics or fringe tracker), a study of the
targets accessible to the VLTI gives a list of 17 Cepheids for which the pulsation will be
measurable at a 3 sigmas level \textit{per measurement}.

The VINCI instrument (Kervella et al. \cite{kervella}), functioning on the same
combination principle as FLUOR (single-mode infrared fibers) is expected to provide an even
higher precision.

One especially interesting target is l~Car (HD84810). Its intrinsic diameter
is evaluated between 180~$D_{\odot}$ (Ripepi et al. 1997) and 195~$D_{\odot}$ (\cite{gieren}).
Its relative proximity (Hipparcos parallax ¹ = 2.16~$\pm$~0.81~mas) gives an expected angular
diameter of more than 3.5~mas. On the 200 meters baseline of the VLTI, l~Car will be fully
resolved, and in the second lobe of its visibility function. It will therefore be possible to
look for features on the photosphere of this star. Since this star is a long period Cepheid
(P=35.5 days), it will be necessary to be particularly careful in the integration of the radial velocity
curve to estimate the distance (see Sect.~\ref{method_amplitude}).

$\zeta$~Gem will also be observable with the VLTI. Based on FLUOR observations, we can expect to
measure the diameter of this star with a statistical precision of about 30~$\mu$as per single
measurement (100 interferograms, or about one minute of observation time). Even considering
calibration issues and the current poor knowledge of the $p$-factor, the final precision on the
distance measurement should be better than 1~\% for 300 observations spread over the
pulsation period.

It is also expected that direct limb darkening measurements will be possible
on six Cepheids (for which visibility is less than 60~\% on the 200 meter
baseline). This will be achieved by observing the star at several baselines and
computing the best fit with three parameters: limb darkening coefficient, mean
angular diameter and amplitude of the pulsational variations.


\section{Conclusion \label{conclusion}}

We have reported the direct measurement of the angular diameter of $\zeta$~Geminorum at a
relative precision of 10~\%, using long-baseline interferometry in the $\rm K^{\prime}$ band.
Our value is consistent with the previous determinations by interferometry (Nordgren et al. \cite{nordgren},
Lane et al. \cite{lane}) and lunar occultation (Ridgway et al. \cite{ridgway}). We have tentatively fit a two
parameter model to our data, based on radial velocity measurements, in order to measure the
amplitude of the pulsation of this Cepheid, but the pulsation could not be detected on a
significant statistical level.

The distance to $\zeta$~Gem was computed using BW linear diameter estimate, and is consistent with the
Hipparcos parallax measurement.
In case infrared interferometric measurements are used together with a visible light
radial velocity curve, any direct distance estimation through the amplitude of the pulsation should
take into account the difference between the pulsational motions of the visible
and infrared photospheres of the star. Otherwise, an underestimation of the true distance could
occur.

The limb darkening factor is still not directly measured but
estimated from a stellar atmosphere model. The use of large ground-based
interferometers will allow to measure it directly. Once direct $p$-factor
measurement and infrared radial velocimetry are available, the distance to the
closest Cepheid stars will be determined in a fully observational way.

\begin{acknowledgements}
P.K. gratefully acknowledges as a graduate student the support of the European Southern
Observatory. This research has made use of the SIMBAD and NASA Astronomical Data Center
databases. The authors would also like to thank the anonymous referee for his
valuable inputs. 
\end{acknowledgements}


\end{document}